# Far-Infrared and submillimeter properties of SDSS galaxies in the *Herschel* ATLAS science demonstration phase field *

Man I Lam[1,2,3], Hong Wu[1,3], Yi-Nan Zhu[1,3] and Zhi-Min Zhou[1,2,3]

[1] National Astronomical Observatories, Chinese Academy of Science, Beijing 100012, China; *linminyi@nao.cas.cn; hwu@bao.ac.cn*
[2] Graduate University of Chinese Academy of Sciences, Beijing 100049, China
[3] Key Laboratory of Optical Astronomy, National Astronomical Observatories, Chinese Academy of Sciences, Beijing 100012, China

**Abstract** Using the *Herschel* ATLAS science demonstration phase data cross-identified with SDSS DR7 spectra, we select 297 galaxies with $F_{250\mu m} > 5\sigma$. The sample galaxies are classified into five morphological types, and more than 40% of the galaxies are peculiar/compact galaxies. The peculiar galaxies show higher far-infrared/submillimeter luminosity-to-mass ratios than the other types. We perform and analyze the correlations of far-infrared/submillimeter and H$\alpha$ luminosities for different morphological types and different spectral types. The Spearman rank coefficient decreases and the scatter increases with the wavelength increasing from 100 μm to 500 μm. We conclude that a single *Herschel* SPIRE band is not good for tracing star formation activities in galaxies. AGNs contribute less to the far-infrared/submillimeter luminosities and do not show a difference from star-forming galaxies. However, the earlier type galaxies present significant deviations from the best fit of star-forming galaxies.

**Key words:** galaxies: formation — galaxies: statistics — galaxies: structure — infrared: galaxies

## 1 INTRODUCTION

After *IRAS* was launched in 1983, infrared astronomy made a breakthrough with the discovery of a type of peculiar object, called an ultra-luminous infrared galaxy (ULIRG) (Sanders & Mirabel 1996). Since that time, more space-based infrared experiments, such as *ISO* and *Spitzer*, have helped us to explore the mid-infrared (MIR) and far-infrared (FIR) emission properties of galaxies (Lonsdale et al. 2006; Soifer et al. 2008). *Herschel* is the first FIR/submillimeter (submm) space telescope which can extend the observations from the FIR to the submm range and reveal properties of cold dust in our universe.

The MIR and FIR emissions of galaxies are usually the indirect tracers of star formation activities, since infrared (IR) emission comes from dust reradiating UV photons emitted by a young stellar population. Most of the stellar light is emitted in the range from the UV to near-IR, with short-lived, massive stars dominating in the UV and more numerous older stars in the near-IR. Dust, produced by the aggregations of metals injected into the ISM by massive stars through stellar winds and supernovae, absorbs the stellar light and re-emits in IR and even submm bands. FIR luminosity with a wavelength less than 200 µm has been demonstrated to be a good star formation tracer by *IRAS* and *ISO* observations (e.g. Hunter et al. 1986; Lehnert & Heckman 1996). In addition, recent studies by *Spitzer* observations indicate that MIR luminosities have good correlations with 1.4 GHz and H$\alpha$ luminosities (e.g. Wu et al. 2005; Calzetti et al. 2007; Zhu et al. 2008; Kennicutt et al. 2009; Calzetti et al. 2010). However, the properties of galaxies in longer wavelengths (such as submm bands) were still unclear before the *Herschel* era. *Herschel* has opened a new 'eye' enabling the study of cool dust in the universe, for example, Domínguez et al. (2012) found 'total' IR luminosity is a good star formation tracer in nearby galaxies by using *Herschel* PACS data. Dunne et al. (2011) found strong dust evolution occurs for massive galaxies at different redshifts, with dust-to-stellar mass ratios being about three to four times larger at $z = 0.4 - 0.5$. The presence of excess submm emission in nearby low-metallicity galaxies shows that these galaxies contain more quantities of cold dust (Dale et al. 2012). *Herschel* also enables better accuracies to model a spectral energy distribution (SED) (e.g. Dale et al. 2012; da Cunha et al. 2012). A large fraction of galaxies in FIR/submm band detections are blue/star-forming galaxies which are obscured by dust (Dariush et al. 2011), but some evidence shows that the old stellar populations may heat the dust grains (Skibba et al. 2011) and contribute to FIR/submm bands in nearby galaxies.

---

* Supported by the National Natural Science Foundation of China.

The morphological classification of galaxies became an important tool for extragalactic research after Hubble's well-known paper was published (Hubble 1926). Many classifiers have been used to compile or extend catalogs from the 20th century that were originally developed to probe the structure of galaxies (e.g. Sandage 1961; de Vaucouleurs et al. 1991). The morphologies of galaxies carry important information which is crucial to our understanding of galaxy formation and evolution (Kennicutt 1998). A number of former investigations show that early-type galaxies exhibit different properties compared to those of late-type ones, such as stellar population, volume of dust and gas and star formation attributes (e.g. Sandage 1986; Davoodi et al. 2006; Shi et al. 2006).

In this paper, we perform a statistical analysis of the FIR/submm properties of *Herschel* galaxies. We describe our sample selection, morphological/spectral classification and luminosity estimation in Sections 2 and 3. The main results and discussions are presented in Section 4, and a summary of this work is given in Section 5. We adopt cosmology constants with $\Omega_m = 0.3$, $\Omega_\Lambda = 0.7$ and $H_0 = 70 \,\mathrm{km\,s^{-1}\,Mpc^{-1}}$ throughout this work.

## 2 SAMPLE

### 2.1 FIR and Submm Sample

The *Herschel* Astrophysical Terahertz Large Area survey (H-ATLAS[1]; Eales et al. 2010) is an open-time key program of the *Herschel* space telescope (Pilbratt et al. 2010), which surveys 550 deg$^2$ of sky with more than 600 hours of observation time. The first released Science Demonstration Phase (SDP) field covers 14 deg$^2$ centered on $9^\mathrm{h}5^\mathrm{m}$, $+0°30'$. The H-ATLAS survey uses both the Photodetector Array Camera and Spectrometer (PACS; Poglitsch et al. 2010) and Spectral and Photometric Imaging REceiver (SPIRE; Griffin et al. 2010) in parallel mode to take images at FIR and submm bands in 100 µm, 160 µm, 250 µm, 350 µm and 500 µm. The archived data obtained from the H-ATLAS SDP catalog (Ibar et al. 2010; Pascale et al. 2011; Rigby et al. 2011) include the object positions (RA and DEC), fluxes from five bands, parameters matched with the Sloan Digital Sky Survey (SDSS; York et al. 2000), *IRAS* and the GAMA project (Smith et al. 2011). There are in total 6876 sources, which are detected by the 250 µm band with a flux greater than $5\sigma$ (32 mJy).

---

[1] *http://www.h-atlas.org/*

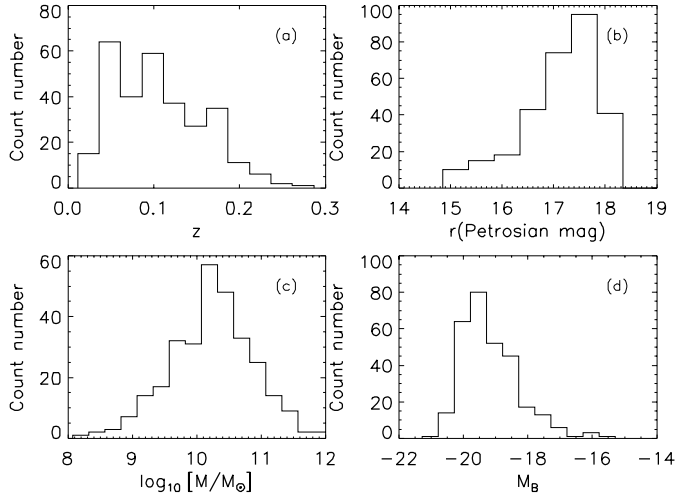

**Fig. 1** Distributions of full sample galaxies. (a) redshift; (b)SDSS $r$-band Petrosian magnitude; (c) stellar masses (details described in Sect. 3.5); (d)$B$-band absolute magnitudes.

## 2.2 SDSS Sample

The optical spectral sample of galaxies is from the cross-identification between the SDSS and H-ATLAS SDP catalogs (Rigby et al. 2011) with a position error of 2.40±0.09″ in radius (Smith et al. 2011). 2423/6876 objects have optical counterparts with the SDSS photometric catalog, but only a small fraction have optical spectra in SDSS. There are finally 301 galaxies which are matched with the MPA emission line catalog of SDSS-DR7[2].

In order to explore the FIR/submm properties in different morphological types of sample galaxies, only galaxies with redshift $z < 0.3$ are chosen as our sample galaxies. This leaves only 298 galaxies in our sample. Since H-ATLAS J090311.6+003906 (SDP.81) is confirmed as a lensing submm galaxy at a high redshift (Lupu et al. 2010; Frayer et al. 2011; Hopwood et al. 2011), we remove it from the sample. The final sample thus includes 297 galaxies. The spectral emission line fluxes are from the MPA catalog.

## 2.3 Sample Distribution

The distributions of redshifts, Petrosian $r$ magnitudes, stellar masses (details are described in Sect. 3.5), and $B$-band absolute magnitudes are shown in Figure 1. Since we set the upper limit of the redshift of sample galaxies to be 0.3, the highest redshift in the sample galaxies is 0.28. Our sample galaxies cover a broad range of stellar masses from $10^8$ to $10^{12}$ $M_\odot$ with a peak above $10^{10}$ $M_\odot$, which is in the range of intermediate mass galaxies. Only a small number (32) of galaxies have an absolute $B$ magnitude ($M_B$) greater than –18 mag, which are defined as dwarf galaxies (Thuan & Martin 1981). Here, $M_B$ is calculated from the SDSS $g$- and $r$-band magnitudes based on the transformation from SDSS magnitude to $UBVI_C R_C$ given by Lupton (2005)[3]. The systematic error of the transformation is 0.011 mag.

---

[2] *http://www.mpa-garching.mpg.de/SDSS/DR7/*
[3] *http://www.sdss.org/dr7/algorithms/sdssUBVRITransform.html*

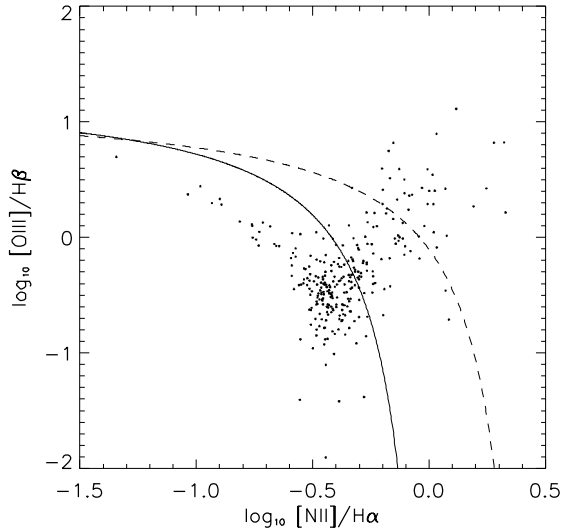

**Fig. 2** BPT diagram: [NII]/Hα versus [OIII]/Hβ. The *solid line* is Kauffmann et al. 2003's line and the *dashed line* is Kewley et al. 2001's line. The galaxies below the *solid line* are defined as star-forming galaxies, those above the *dashed line* are AGNs, and the others between the two lines are classified as the composite galaxies.

## 3 CLASSIFICATIONS, LUMINOSITIES AND STELLAR MASSES

### 3.1 Spectral Classification

The optical spectral classification of galaxies adopts the traditional BPT diagram: [NII]/Hα versus [OIII]/Hβ (Baldwin et al. 1981; Veilleux & Osterbrock 1987), as shown in Figure 2. The lower line is from Kauffmann et al. (2003, hereafter Ka03), and the upper line is from Kewley et al. (2001, hereafter Ke01). The galaxies in which line ratios are lower than Ka03 are defined as star-forming galaxies, and those at positions above Ke01 are classified as AGNs. The galaxies between the two lines are classified as composite galaxies, which represents the co-existence of star-formation and a central AGN in galaxies (Kewley et al. 2006). Composite galaxies are similar to the mixture-type galaxies given by Wu et al. (1998), based on three BPT diagrams in Veilleux & Osterbrock (1987). If a galaxy is shown as a star-forming case in one diagram but as an AGN in another, then it is classified as a mixture-type galaxy (Wu et al. 1998).

The numbers of star-forming, AGN and composite galaxies are shown in Table 1. About 67% (199/297) of sample galaxies and 74% of galaxies with $EW(H\alpha) > 5$ are classified as star-forming galaxies. 20% (60/297) are composite galaxies and the other 12% (35/297) are AGNs.

**Table 1** Number of Sample Galaxies with Different Spectral Types

| Sample | Star-forming | Composite | AGN | Total |
|---|---|---|---|---|
| Whole | 199 | 60 | 35 | 297 |
| $EW(H\alpha) > 5$ | 181 | 45 | 15 | 243 |
| $EW(H\alpha) < 5$ | 18 | 15 | 20 | 54 |

**Table 2** Number of Different Morphological Types of Sample Galaxies in Five *Herschel* Bands

| Band (1) | Edge (2) | E/S0 (3) | Early-type Spiral (4) | Late-type Spiral (5) | Pec (6) | Compact (7) | Total (8) |
|---|---|---|---|---|---|---|---|
| 100 μm | 7 | 5 | 9 | 15 | 25 | 5 | 68 |
| 160 μm | 12 | 7 | 17 | 25 | 36 | 7 | 106 |
| 250 μm | 36 | 23 | 58 | 55 | 100 | 23 | 297 |
| 350 μm | 36 | 23 | 58 | 55 | 100 | 23 | 297 |
| 500 μm | 31 | 21 | 55 | 51 | 90 | 19 | 269 |
| All Bands | 6 | 5 | 8 | 11 | 25 | 4 | 59 |

Column (1): *Herschel* Bands; Col. (2): edge-on galaxies; Col. (3): elliptical galaxies and lenticular galaxies; Col. (4): spiral galaxies with earlier than Sb morphologies; Col. (5): spiral galaxies with later than Sb morphologies; Col. (6): peculiar galaxies; Col. (7): compact galaxies; Col. (8): total number.

### 3.2 Morphological Classification

The visual morphological classification of sample galaxies is based on the criterion of the Third Reference Catalogue of bright galaxies (RC3; de Vaucouleurs et al. 1991 according to the bulge ratio, spiral arm and interaction features). All sample galaxies are divided into five different morphological types: E/S0, early-type spiral (earlier than Sb), late-type spiral (later than Sb), peculiar and compact. The peculiar type signifies the galaxies with asymmetric morphologies. The visual classification is performed based on $r$-band images by three of the authors, and most of the classifications are consistent. For those inconsistent cases, we adopt a consensus decision after discussion.

Only two galaxies are unclassified, because of their larger redshift and unclear features. The galaxies with a large inclination angle of $i > 60°$ are labeled as edge-on galaxies (generally defined as $i = 90°$) without further classification. In this case, 12% of the sample galaxies are classified as edge-on galaxies. We do not identify the bar feature, since it is difficult to identify in our sample galaxies. The statistical results of morphological types are shown in Table 2.

### 3.3 FIR/submm Luminosities

Figure 3 shows the distributions of FIR/submm luminosity in five bands. The K-corrections of five FIR/submm bands are based on the best-fitting SED model of nearby starburst galaxy M82 (Siebenmorgen & Krügel 2007). The monochromatic 250 μm luminosities of our sample galaxies are between $10^8$ to $10^{11}$ $L_\odot$. The peak of distribution shifts to lower luminosity from 100 μm to 500 μm. Figure 4 shows the comparison of 100 μm luminosities of galaxies drawn from *IRAS* and *Herschel*. Both 100 μm luminosities are in good agreement except for two that deviate at the luminous end. The agreement confirms the flux calibration of the *Herschel* bands. Two galaxies that deviate show higher *IRAS* 100 μm luminosities, which may due to a lower spatial resolution and larger error bars of *IRAS*.

### 3.4 Hα Luminosity

Since the SDSS spectra are taken by a 3″ diameter fiber, the Hα fluxes of sample galaxies (especially low redshift galaxies) are only from the central region. Aperture corrections are required. To obtain the Hα luminosities of whole galaxies, we adopt the equation (A2) of Hopkins et al. (2003), which is shown below

$$L_{\mathrm{H}\alpha} = 4\pi D_L^2 S_{\mathrm{H}\alpha} \times 10^{-0.4(r_{\mathrm{Petro}} - r_{\mathrm{fiber}})} . \tag{1}$$

Here, $D_L$ is the luminosity distance; $S_{\mathrm{H}\alpha}$ is flux of Hα emission; $r_{\mathrm{Petro}}$ and $r_{\mathrm{fiber}}$ are the SDSS $r$-band Petrosian and fiber magnitudes, respectively.

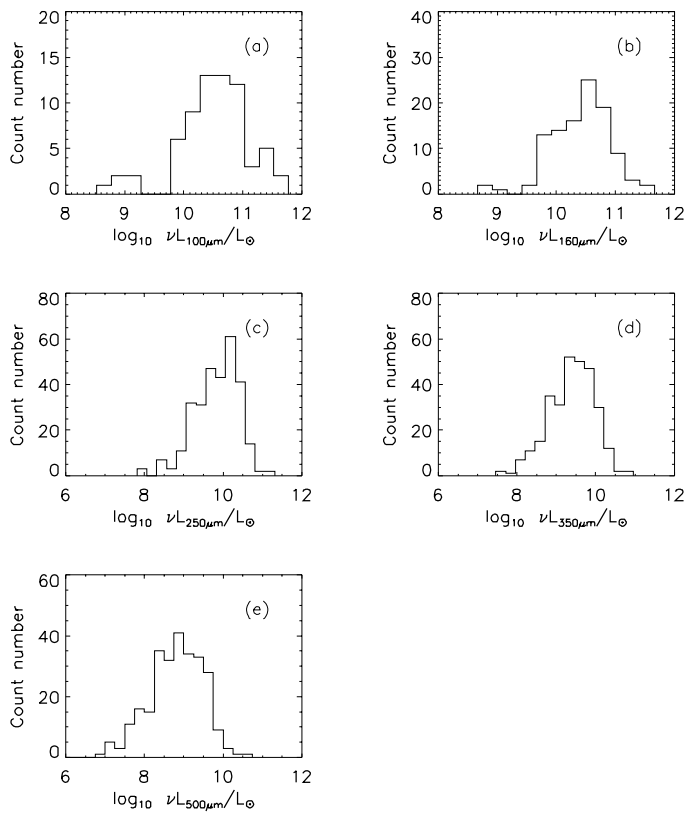

**Fig. 3** FIR/submm luminosity distributions of five H-ATLAS bands. (a) 100 μm (PACS); (b) 160 μm (PACS); (c) 250 μm (SPIRE); (d) 350 μm (SPIRE); (e) 500 μm (SPIRE).

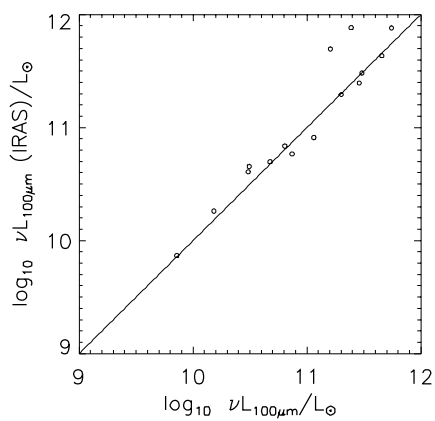

**Fig. 4** *Herschel* 100 μm luminosity versus *IRAS* 100 μm luminosity. The *solid line* shows equal values.

We consider both extinction corrections from the Milky Way and galaxies themselves. First, we adopt the Milk Way extinction curve from Cardelli et al. (1989) to do the Milky Way correction with $R_V = 3.1$. Then the intrinsic extinction correction is performed by using Calzetti (2001)'s model. The $E(B-V)$ value is obtained from the Balmer decrement $F_{\mathrm{H}\alpha}/F_{\mathrm{H}\beta}$.

### 3.5 Stellar Masses

The stellar masses of our sample galaxies are calculated by using the formula from Bell et al. (2003)

$$\log\left(\frac{M_{\mathrm{star}}}{M_\odot}\right) = -0.4(M_{r,AB} - 4.67) + a_r + b_r(g-r)_{AB} + 0.15\,, \quad (2)$$

where $M_{r,AB}$ is the absolute $r$ magnitude and $(g-r)_{AB}$ is the rest-frame color. The term 4.67 is the absolute $r$ band magnitude of the Sun. All magnitudes are in the AB magnitude system. The photometric K-correction adopts the code from Blanton & Roweis (2007) (kcorrect). The coefficients $a_r$ and $b_r$ come from table 7 of Bell et al. (2003). The stellar mass calculated by Equation (3) has an uncertainty of about 0.3 dex due to the effects of reddening, galaxy age and star formation history.

## 4 RESULTS AND DISCUSSIONS

### 4.1 Morphologies

The fraction of morphological types in our sample is E/S0: S(Sa-Sd): Peculiar: Compact = 0.08: 0.38: 0.34: 0.08. Compared with E(E/S0-S0): S(S0a-Sdm): Im = 0.40: 0.57: 0.014 of an SDSS optically selected sample (Fukugita et al. 2007), our sample is obviously inclined toward later spiral and peculiar cases, as many of the infrared selected galaxies are linked to interacting systems. Similar to Fukugita et al. (2007), Li et al. (2007) obtain a morphological fraction of E/S0: S(Sa-Sd): Irr = 0.47: 0.51: 0.014, which is also different from this work. Though Li et al. (2007)'s sample is matched from *Spitzer* MIR and SDSS optical samples, it is almost completely an optically selected sample, instead of being IR selected, since the *Spitzer* MIR observation is much deeper than that of the optical cut-off of $r = 15.9$ mag. Our samples are selected from 250 μm with an optical flux cut-off of $r = 17.8$ mag. They still represent an IR sample, because only a small fraction of SDSS galaxies with $r < 17.8$ mag are detected in the 250 μm band.

Wang et al. (2006) select a sample of luminous IR galaxies with $\log L_{\mathrm{IR}} > 10^{11} L_\odot$ from Cao et al. (2006)'s catalog, and the fraction of peculiar/compact galaxies increases from 40% ($11 < \log L_{\mathrm{IR}}/L_\odot < 11.3$) to 100% ($\log L_{\mathrm{IR}}/L_\odot > 11.6$). Wu et al. (1998) obtain a fraction of 56% for the interacting/merging systems based on a sample of more luminous IR galaxies ($\log L_{\mathrm{IR}}/L_\odot > 11.5$). Zou et al. (1991) construct a sample of ULIRGs ($\log L_{\mathrm{IR}}/L_{\mathrm{sun}} > 12$), and found that $61\pm12\%$ were in interacting/merging systems. Our sample contains a smaller fraction of peculiar galaxies than those of Zou et al. (1991) and Wu et al. (1998), and a more luminous sub-sample than that of Wang et al. (2006), but is similar to the less luminous sample ($11 < \log L_{\mathrm{IR}}/L_\odot < 11.3$) of Wang et al. (2006). It can be explained by Figure 3, which shows that most of our sample selected from 250 μm has 100 μm luminosities much lower than $10^{12} L_\odot$ and peaks around $10^{10.5} L_\odot$. Rowlands et al. (2012) select galaxies from H-ATLAS/GAMA and obtain the morphological fraction of early-type: late-type: Merger: Unknown = 0.055: 0.282: 0.25: 0.206; however their sample contains a large fraction of unknown galaxies, because their deeper $r$ magnitude cut-off is 19.4.

### 4.2 Luminosity-to-Mass Ratios

The distributions of FIR/submm luminosity-to-mass ratios are shown in Figure 5. The different morphological types are shown as different line styles. From the figure, we find that the median

186 M. I Lam et al.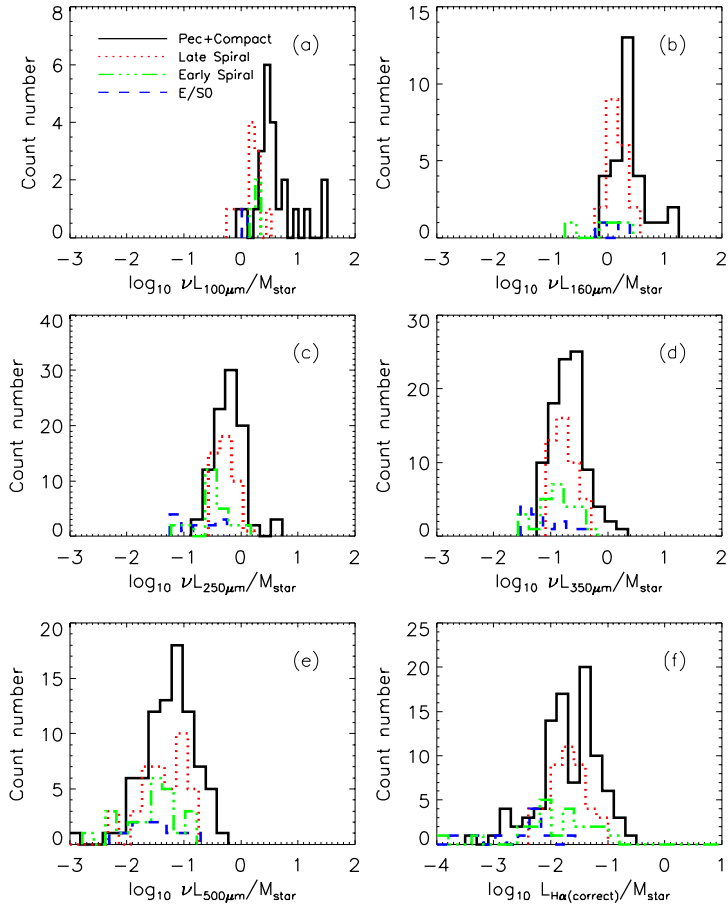

**Fig. 5** The distributions of luminosity-to-mass ratios (FIR/submm & H$\alpha$) for different morphological types. *Solid line*: peculiar and compact galaxies; *dotted line*: late-type spiral galaxies; *dash-dotted line*: early-type spiral galaxies; *dashed line*: E/S0 galaxies.

FIR/submm luminosity-to-mass ratios decrease from 100 μm to 500 μm (Table 3). For instance, the median value of peculiar/compact galaxies is 0.51 at 100 μm, and decreases to –1.20 at 500 μm. Though many of the galaxies are not detected in PACS bands and there is some selection effect, the trend still remains. Because the peak of FIR emission from dust is around 100 μm in the star formation region, the largest FIR/submm luminosity-to-mass ratio appears in 100 μm and then decreases as wavelength increases. Compared with other types, the peculiar/compact galaxies show the highest FIR/submm luminosity-to-mass ratio. As Figure 5(a)–(d) displays, the peculiar/compact galaxies peak at a higher $\nu L\nu/M_{\rm star}$ than other types. It indicates a trend from peculiar/compact galaxies to early-type galaxies, implying that the peculiar/compact galaxies may host more active star formation processes than others, which is enhanced by galaxy interaction.

In Figure 5(f) and Table 3, $L_{\rm H\alpha}/M_{\rm star}$ ratios of different morphological types show a similar trend to the FIR/submm ratios. Peculiar/compact galaxies present the highest median H$\alpha$ luminosity-to-mass ratio of –1.68, and late-type spiral galaxies have the a slightly lower value of –1.72. Early-

**Table 3** Median of Luminosity-to-Mass Ratio for Different Morphological Types

| Galaxy Type | $\log \frac{\nu L_{100\mu m}}{M_{star}}$ | $\log \frac{\nu L_{160\mu m}}{M_{star}}$ | $\log \frac{\nu L_{250\mu m}}{M_{star}}$ | $\log \frac{\nu L_{350\mu m}}{M_{star}}$ | $\log \frac{\nu L_{500\mu m}}{M_{star}}$ | $\log \frac{L_{H\alpha}}{M_{star}}$ | $\log sSFR_{H\alpha}$ |
|---|---|---|---|---|---|---|---|
| (1) | (2) | (3) | (4) | (5) | (6) | (7) | (8) |
| Peculiar/Compact | 0.51 | 0.38 | −0.22 | −0.67 | −1.20 | −1.68 | −9.20 |
| Late-type Spiral | 0.19 | 0.16 | −0.25 | −0.79 | −1.38 | −1.72 | −9.24 |
| Early-type Spiral | 0.30 | 0.16 | −0.45 | −0.82 | −1.50 | −1.90 | −9.42 |
| E/S0 | 0.018 | 0.21 | −0.71 | −1.18 | −1.55 | −2.36 | −9.88 |

Column (1): morphological Hubble types. An early-type spiral means that the morphological type is earlier than Sb; Cols. (2)–(7): luminosity-to-mass ratios of five FIR/submm bands and H$\alpha$, in units of $\log L_\odot/M_\odot$; Col. (8): sSFR refers to the specific star formation rate, SFR/$M_{star}$ with H$\alpha$ fluxes by using Kennicutt (1998), in units of $\log (yr^{-1})$.

type spirals and E/S0 show lower ratios of –1.90 and –2.36, respectively. Since single submm luminosities may not be a good SFR tracer and may be related to another process, we used $L_{H\alpha}/M_{star}$ to represent the specific star formation rate (sSFR), which describes the intensity of star formation. To derive the sSFR, we transformed $L_{H\alpha}$ to the SFR by using the Kennicutt (1998) formula

$$\mathrm{SFR}(M_\odot \, \mathrm{yr}^{-1}) = 7.9 \times 10^{-42} L_{H\alpha}(\mathrm{erg \, s}^{-1}). \tag{3}$$

The median sSFRs obtained for peculiar/compact, late-type spirals, early-type spirals and E/S0 galaxies are –9.20, –9.24, –9.42 and –9.88, respectively. These are much higher than the values of –9.52, –9.90 and –10.15 for dwarf galaxies, spirals, and early types respectively from the KINGFISH sample (table 2, Skibba et al. 2011), which is a nearby optically selected sample with lower FIR/submm luminosities (Kennicutt et al. 2011). The sSFRs in our sample were consistent with the GAMA/H-ATLAS result, based on Wijesinghe et al. (2011), and showed a broad range of sSFR from –12.0 to –8.8. Compared to the KINGFISH sample, the early-type galaxies in our sample were inclined to higher sSFR, which indicates that our selection criteria made it easy to select those early-type dust rich galaxies, and they appeared in a narrower range of $L(\mathrm{FIR})/M_{star}$ than that of $L(H\alpha)/M_{star}$ in our sample.

Comparing (a)–(d) and (f) in Figure 5, all galaxies in our sample show a similar FIR/submm distribution but very different H$\alpha$ emission. The different ratio widths between H$\alpha$ and FIR/submm may reveal that FIR/submm and H$\alpha$ do not originate in the same region. The H$\alpha$ emission is coming from regions where massive stars recently formed but the FIR/submm is emitted by the cold dust, which may be comnnected to the old stellar population or evolved stars in the entire galaxy. A recent study based on nearby galaxies shows that the FIR/submm emissions in SPIRE bands in those galaxies are coming from whole galaxies and have a good correlation with 1.6 µm (Bendo et al. 2012). Note that the aperture correction may affect our results if stars were forming at the center of a galaxy or stars were forming in the disk.

### 4.3 Correlations of FIR/submm Luminosities

FIR luminosity has proven to be a good star formation tracer from previous *IRAS*, *ISO* and *Spitzer* observations.

Figure 6 shows the correlations between five FIR/submm luminosities. Due to the relatively lower sensitivities of *Herschel* PACS 100 µm, only ∼30 star-forming galaxies have been detected in this band. In the figure, the best fitting parameters tend to change with wavelength. From 160 µm to 500 µm, the Spearman rank coefficient decreases from 0.98 to 0.88 and the scatter increases from 0.09 dex to 0.25 dex. The trend of slopes and scatters from 100 µm to 500 µm indicates that a young stellar population may not be the only contributor to the flux in the longer wavelength (submm) band, if we assume that it dominates 100 µm. There may exist some other contributors.

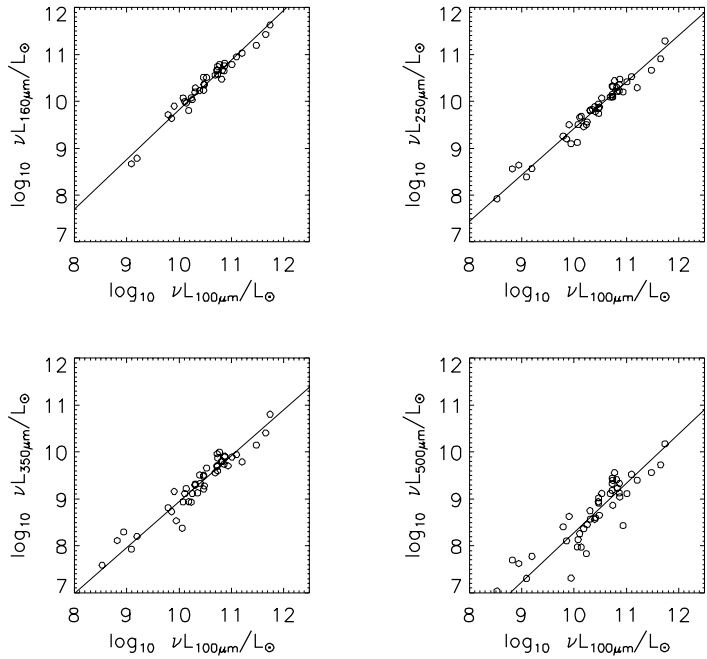

**Fig. 6** Correlations among *Herschel* five FIR/submm bands. With the wavelength increasing, the correlation coefficient decreases and the scatter increases. We only plot 100 μm vs. 160 μm, 100 μm vs. 250 μm, and 100 μm vs. 500 μm. 100 μm vs. 100 μm must be a linear correlation, so we omitted this plot.

### 4.4 Correlations between FIR/submm and Hα luminosities

It has been shown in the *IRAS* and *Spitzer* era that the Hα and MIR/FIR continuums are tightly correlated (e.g. Hunter et al. 1986; Lehnert & Heckman 1996; Roussel et al. 2001; Wu et al. 2005; Calzetti et al. 2005). They are usually used as star formation tracers with the assumption that Hα emission is from a young stellar population and the IR continuum is from UV photons that are absorbed and re-emitted by dust (Calzetti et al. 2010). However, the tightness of the correlation between Hα and submm is still unclear. We try to explore whether there is a tight correlation. Before that, we must consider the possible factors, which could affect our results.

Signal-to-Noise (S/N) To consider the possible effect from low S/N, we exclude the sources with low signal-to-noise and only select those with S/N> 10 in each band to reduce the effect from random noise. Although the scatter in the submm band tends to be a little smaller, the fitted results have a similar trend and do not show a significant difference from those in Table 4.

Morphologies Figure 7 shows the FIR/submm-to-Hα correlations. We mark the morphologically classified E/S0 galaxies as red pluses. It is shown that the E/S0 galaxies deviate from the best fitting lines in all the three submm bands although the scatter is larger. To confirm such a trend, we also classify the early-type by Hα equivalent width (EW(Hα)) from the optical spectra. The sample galaxies are divided into two types by EW(Hα)=5. There are 53 galaxies with EW(Hα)<5 that are classified as early-type. In addition, 43/53 are morphologically classified as E/S0, early-type spiral, or edge-on galaxies. Similarly, the galaxies with EW(Hα)<5 also deviate from the best fitting lines and are biased to the same side as E/S0 galaxies. Both classifications show that early-type galaxies do not follow the general relationship of star-forming galaxies and the FIR/submm-to-H$\alpha$ correlations.

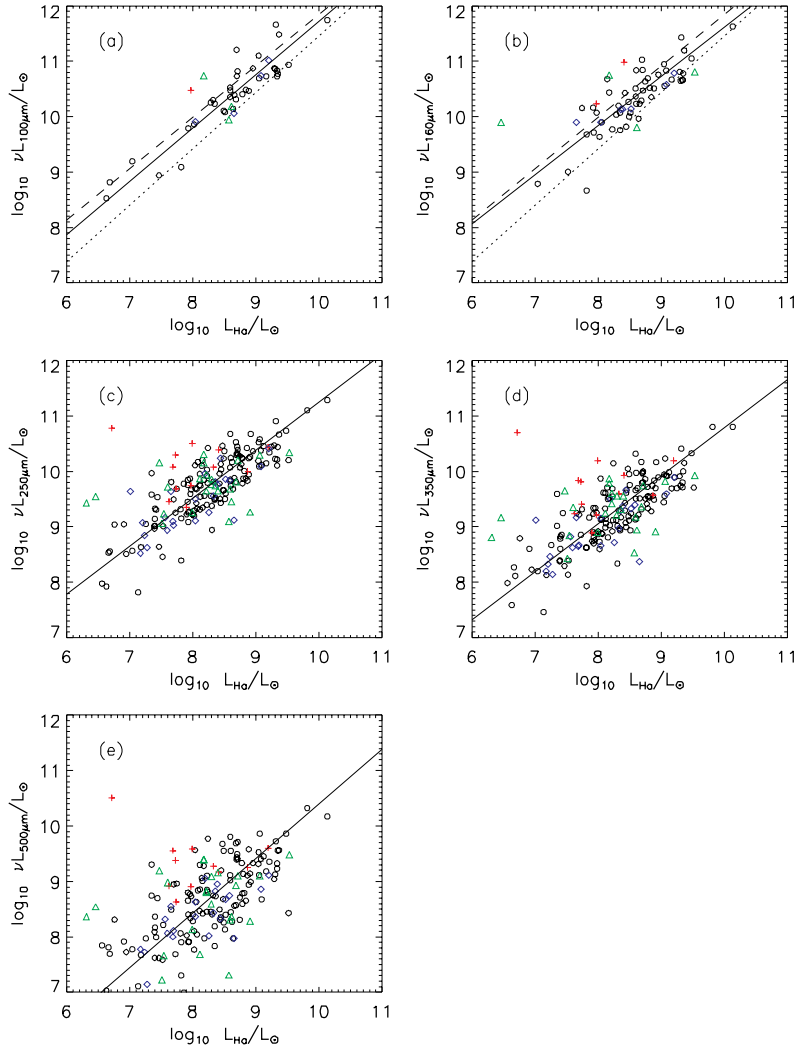

**Fig. 7** Correlations between FIR/submm and H$\alpha$ luminosities of star-forming galaxies. The *red-pluses* are E/S0 galaxies and *blue-diamonds* are edge-on galaxies. The best fittings are shown as *solid lines*. The *dotted* and *dashed lines* are the fittings in 70 μm and 160 μm from Zhu et al. (2008), respectively.

**AGN Contributions**  We test how the AGNs affect the correlations and we plot the galaxies with EW(H$\alpha$) >5 in Figure 8 and label star-forming galaxies, AGNs and composite galaxies in different symbols. The best fittings of star-forming galaxies are shown as lines in the figure. The correlation results in different bands are listed in Table 4. Low luminosity AGNs and composite galaxies do not influence the fitting parameters shown in the table, but a small fraction of Seyfert 2 galaxies have higher H$\alpha$ emissions in our galaxy sample.

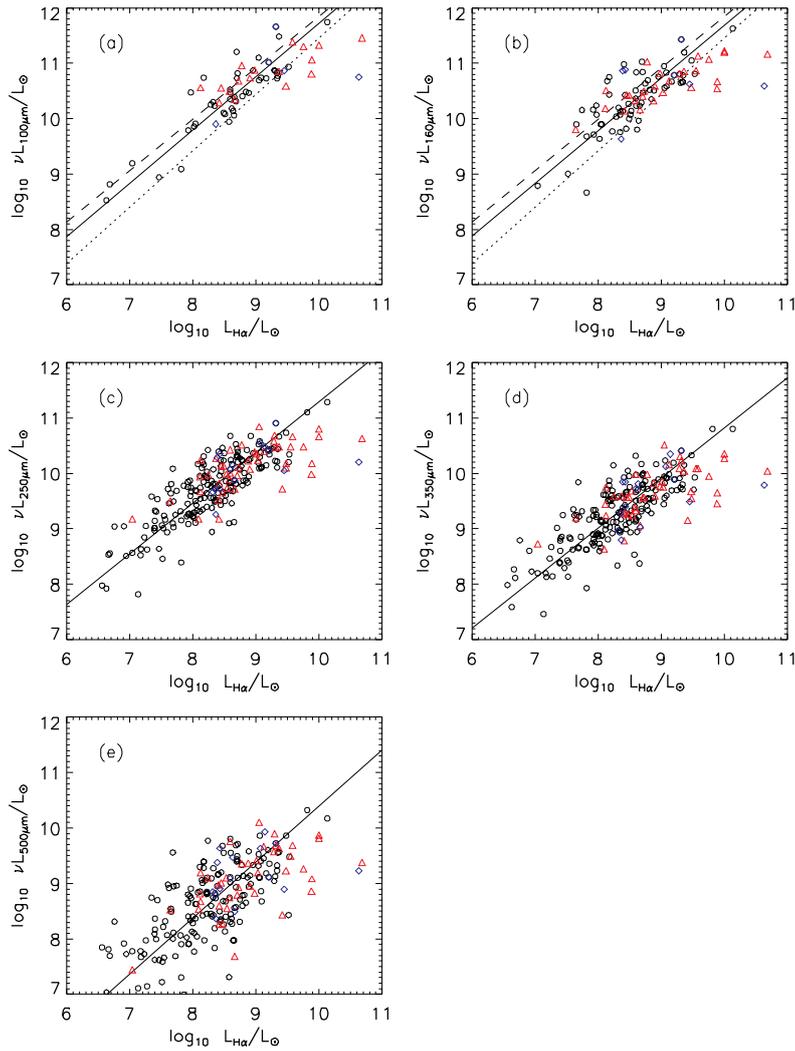

**Fig. 8** The correlations between FIR/submm and Hα luminosities of sample galaxies with EW(Hα) > 5. *Circles*: starburst galaxies; *Triangles*: composite galaxies; *Diamonds*: AGNs; *Solid line*: best fitting of starburst galaxies. The *dotted* and *dashed lines* are fitting lines in 70 μm and 160 μm from Zhu et al. (2008) respectively.

Contamination from Stars  Because of the lower resolution of *Herschel* sources, we should consider the possible contamination from nearby foreground stars or background galaxies seen in the optical band. After checking all galaxies in our sample, only a small number of galaxies have companion stars within $1\sigma$ that do not show any deviation from the others. Since these stars are located at a high galactic latitude region, their contribution to FIR/submm can be neglected.

Considering the above aspects, we present the correlations between FIR/submm luminosities and Hα luminosity for the whole region of star-forming galaxies and galaxies with EW(Hα)>5 in Figures 7 and 8, respectively. Using the bivariate regression, we obtain the best fitting [$\log_{10}(y) =$ $a + b \log_{10}(x)]$ for whole star-forming galaxies, star-forming galaxies with EW(H$\alpha$) > 5 and galaxies with EW(H$\alpha$) >5. The best fitting results in each band are shown in Table 4. Here we only show the best fitting for star-forming galaxies with EW(H$\alpha$)>5 as follows:

$$\log_{10} \nu L_{100\mu m} = (2.09 \pm 0.19) + (0.96 \pm 0.06) \times \log_{10} L_{H\alpha}, \qquad (4)$$

$$\log_{10} \nu L_{160\mu m} = (2.18 \pm 0.23) + (0.95 \pm 0.08) \times \log_{10} L_{H\alpha}, \qquad (5)$$

$$\log_{10} \nu L_{250\mu m} = (2.14 \pm 0.11) + (0.92 \pm 0.04) \times \log_{10} L_{H\alpha}, \qquad (6)$$

$$\log_{10} \nu L_{350\mu m} = (1.78 \pm 0.12) + (0.90 \pm 0.04) \times \log_{10} L_{H\alpha}, \qquad (7)$$

$$\log_{10} \nu L_{500\mu m} = (0.32 \pm 0.18) + (1.01 \pm 0.06) \times \log_{10} L_{H\alpha}. \qquad (8)$$

Compared with the fitting of star-forming galaxies in previous studies, we find that the scatters are obviously larger than those in MIR (Wu et al. 2005; Zhu et al. 2008). In both Figures 7 and 8, we also plot the lines from Zhu et al. (2008) based on samples of star-forming galaxies from *Spitzer* fields. Our fittings in *Herschel* 100 μm and 160 μm agree well with those in *Spitzer* 70 μm (dotted lines) and 160 μm (dashed lines) of Zhu et al. (2008). The 100 μm fit is just between *Spitzer* 70 μm and 160 μm and the 160 μm fit only shifts a little from the *Spitzer* 160 μm line.

Among the fittings in different *Herschel* bands, the tightest is $\nu L_{100\mu m} - L_{H\alpha}$ with the least scatter and the largest Spearman rank coefficient in each of the three samples of galaxies. The result supports that the origin of *Herschel* 100 μm luminosity is from star-formation. However in the longer wavelength band, the correlation tends to be weaker. For star-forming galaxies with EW(H$\alpha$)>5 as an example, the Spearman rank coefficient changes from 0.85 for 100 μm to 0.68 for 500 μm with scatters from 0.22 dex for 100 μm to 0.39 dex for 500 μm. Such a trend may be due to the distribution of temperatures in cold dust. Hwang et al. (2010) find the galaxies at $z > 0.5$ appear to have lower temperatures compared to their control sample galaxies with similar luminosities in the local universe. Another possibility is that star formation is not the only contributor to the luminosity in the SPIRE bands, and other processes may also heat the cold dust grains.

The interstellar medium (ISM) is ubiquitous in galaxies, and is diffuse, such as the cirrus in the Milky Way discovered by IRAS. The cold components of the ISM could mainly be heated by old stars or evolved stars, and thus the cold dust emission in a single submm may not be used in determining SFRs (Gordon et al. 2004). Rowlands et al. (2012) show that FIR/submm luminosities can be contributed by ISM, and its fraction can even be more than 50% in spiral galaxies. Therefore the ISM can contribute partly to FIR/submm luminosities of galaxies even if there is no ongoing star formation occurring. Since in such a case star formation does not dominate the submm bands anymore, the single submm and H$\alpha$ bands lose their common physical link and the tight correlation broadens. Single submm bands are problematic as a star formation tracer, even though 100 μm luminosity has still proven to be a good star formation tracer, as previously noted (Lonsdale Persson & Helou 1987; Sauvage & Thuan 1992). The increased dispersion at the longer wavelengths may be partly due to the change in the properties of the cool dust, such as emissivity changes in grain size or diffuse dust components associated with the young stellar population (Popescu et al. 2011).

**4.5 FIR/submm Emission of the Early-Type Galaxies**

Generally, there are almost no star formation activities in early-type galaxies. Therefore, their FIR/submm luminosities would be much lower than those in star-forming galaxies, which is the reason for a lower fraction of early-type galaxies in our sample. However, there are tens of early-type galaxies in our sample. The detection of early-type galaxies by *IRAS* reveals that the FIR emission in E/S0 galaxies was independent of apparent magnitude or redshift of the system (Goudfrooij & de Jong 1995) and they are detected in about half of the systems (Knapp et al. 1989). To explain their relatively higher FIR/submm luminosities from the previous section, we have shown FIR/submm-to-H$\alpha$ correlations in Figure 7. The early-type galaxies seem biased toward the upper side of the correlation in the submm bands, which indicates that they present relatively higher submm luminosity compared with H$\alpha$ luminosity. Half of the early-type galaxies in our sample are galaxies in the star forming sequence of galaxies, but the rest are main sequence galaxies. Such submm luminosity should be related not only to the star forming region in the main sequence early-type galaxies (Brinchmann et al. 2004; Noeske et al. 2007; Peng et al. 2010).

A cool, cirrus-dominated emission has usually been attributed to dust heating from the general stellar radiation field, including the visible radiation from old stars (Lonsdale Persson & Helou 1987; Sauvage & Thuan 1992; Rowan-Robinson & Crawford 1989; Walterbos & Greenawalt 1996; Kennicutt 1998). Although young stars are still the main sources of FIR/submm luminosity in some early-type galaxies, such as barred galaxies with strong nuclear starbursts and some unusually blue objects (e.g. Zhou et al. 2011; Kormendy & Fisher 2005; Kennicutt 1998), many early-type galaxies show no independent evidence of high FIR/submm emission. These unusual early-type galaxies suggested that the older stars are responsible for much of the FIR/submm emission. Nearby galaxies like M81 revealed that about 100% of the >100 μm emission originated from dust heated by the evolved disk and bulge stars (Bendo et al. 2010, 2012). Rowlands et al. (2012) showed that more than 70% of luminosities of early-type galaxies can be contributed by ISM. Smith et al. (2012) obtained similar results based on an optically selected sample in the *Herschel* Reference Survey, and showed that the ISM can be heated by the old stellar population and re-emitted by FIR/submm. However, to produce such high FIR/submm luminosity, the early-type galaxies must contain a large amount of dust, comparable to spirals (Smith et al. 2012), but neither AGB stars nor supernovae can provide enough dust mass (Rowlands et al. 2012). Therefore the mass and origin of dust in early-type galaxies are still problems to be solved and explained in future observations and theoretical models.

The H$\alpha$ emission is measured in many of our early-type galaxies, although the galaxies look quiescent and do not show disturbed features in their optical images. One possibility is that they are

**Table 4** Correlation Parameters of Sample Galaxies

| Sample | $y$ | $x$ | $a$ | $b$ | $s$ | $r$ | $c$ | $N$ |
|---|---|---|---|---|---|---|---|---|
| (1) | (2) | (3) | (4) | (5) | (6) | (7) | (8) | (9) |
| whole star-forming | $\nu L_{100\mu m}$ | $L_{H\alpha}$ | $2.09 \pm 0.19$ | $0.96 \pm 0.06$ | 0.22 | 0.85 | $1.79 \pm 0.26$ | 45 |
| | $\nu L_{160\mu m}$ | $L_{H\alpha}$ | $2.74 \pm 0.21$ | $0.89 \pm 0.07$ | 0.24 | 0.77 | $1.79 \pm 0.30$ | 64 |
| | $\nu L_{250\mu m}$ | $L_{H\alpha}$ | $2.59 \pm 0.12$ | $0.87 \pm 0.04$ | 0.27 | 0.73 | $1.46 \pm 0.38$ | 199 |
| | $\nu L_{350\mu m}$ | $L_{H\alpha}$ | $2.14 \pm 0.13$ | $0.87 \pm 0.04$ | 0.29 | 0.70 | $1.01 \pm 0.41$ | 199 |
| | $\nu L_{500\mu m}$ | $L_{H\alpha}$ | $0.55 \pm 0.19$ | $0.98 \pm 0.06$ | 0.42 | 0.59 | $0.40 \pm 0.53$ | 182 |
| EW(H$\alpha$) > 5 (only star-forming) | $\nu L_{100\mu m}$ | $L_{H\alpha}$ | $2.09 \pm 0.19$ | $0.96 \pm 0.06$ | 0.22 | 0.85 | $1.79 \pm 0.26$ | 45 |
| | $\nu L_{160\mu m}$ | $L_{H\alpha}$ | $2.18 \pm 0.23$ | $0.95 \pm 0.08$ | 0.24 | 0.76 | $1.76 \pm 0.27$ | 63 |
| | $\nu L_{250\mu m}$ | $L_{H\alpha}$ | $2.14 \pm 0.11$ | $0.92 \pm 0.04$ | 0.25 | 0.81 | $1.44 \pm 0.30$ | 181 |
| | $\nu L_{350\mu m}$ | $L_{H\alpha}$ | $1.78 \pm 0.12$ | $0.90 \pm 0.04$ | 0.26 | 0.78 | $0.99 \pm 0.32$ | 181 |
| | $\nu L_{500\mu m}$ | $L_{H\alpha}$ | $0.32 \pm 0.18$ | $1.01 \pm 0.06$ | 0.39 | 0.68 | $0.37 \pm 0.44$ | 67 |
| EW(H$\alpha$) > 5 (including AGN & Composite) | $\nu L_{100\mu m}$ | $L_{H\alpha}$ | $3.36 \pm 0.16$ | $0.81 \pm 0.05$ | 0.23 | 0.84 | $1.73 \pm 0.31$ | 66 |
| | $\nu L_{160\mu m}$ | $L_{H\alpha}$ | $3.74 \pm 0.16$ | $0.77 \pm 0.05$ | 0.25 | 0.74 | $1.70 \pm 0.34$ | 96 |
| | $\nu L_{250\mu m}$ | $L_{H\alpha}$ | $2.76 \pm 0.10$ | $0.84 \pm 0.03$ | 0.25 | 0.80 | $1.39 \pm 0.32$ | 241 |
| | $\nu L_{350\mu m}$ | $L_{H\alpha}$ | $2.33 \pm 0.11$ | $0.83 \pm 0.04$ | 0.27 | 0.77 | $0.94 \pm 0.34$ | 241 |
| | $\nu L_{500\mu m}$ | $L_{H\alpha}$ | $0.90 \pm 0.15$ | $0.93 \pm 0.05$ | 0.39 | 0.67 | $0.32 \pm 0.46$ | 223 |

Column (1): samples used for correlation analysis; Cols. (2) and (3): luminosities of the $y$ and $x$ axes; Cols. (4) and (5): the nonlinear fitting coefficients $a$ and $b$: $\log_{10} = a + b \log_{10}(x)$; Col. (6): the standard deviation $s$ of the fitting; Col. (7): the Spearman rank coefficient $r$; Col. (8): the linear fitting coefficient $c$: $\log_{10} = c + \log_{10}(x)$; Col. (9): the number of sample galaxies used for fitting.

undergoing minor mergers, which invoke gas infalling to the galactic center, that trigger starbursts over a small scale. It is a common phenomenon in our universe (Zaritsky et al. 1997; Ostriker & Tremaine 1975), since it is easier for early-type galaxies to attract smaller galaxies (Haynes et al. 2000), because of their high stellar mass and large gravitational potential. Another possibility is that some early-type galaxies are post-starburst galaxies. Post-starburst galaxies can appear faint in H$\alpha$ emission since their star formation was shut down more than 10 Myr ago, but FIR can remain strong on a scale of 100 Myr after shut down. The existence of galaxies which show a very faint $\alpha$ emission with strong Balmer absorptions (e.g H$\beta$, H$\gamma$) but without any [OII] emission in our sample can be classified as E+A galaxies (Goto 2004; Goto et al. 2008). Yamauchi & Goto (2005) suggested that early-type E+A galaxies (or K+A galaxies) are post-starburst galaxies that have undergone centralized starbursts arising from galaxy-galaxy interactions and mergers. In both cases, the central star formation can explain some of the FIR/submm luminosity in early-type galaxies.

Considering the density of FIR/submm sources, it is possible that an optical early-type galaxy is associated with a coincident background FIR/submm source. The background source will penetrate the foreground because no galaxy is optically thick in FIR/submm on a large scale. Since the early-type galaxy has a large gravitational potential, it could even act as a gravitational lens system and enhance the emission from background sources. *Herschel* has found a lot of lensed background FIR/submm galaxies (e.g. Rawle et al. 2010; Rex et al. 2010; Ivison et al. 2010; Negrello et al. 2010; Cox et al. 2011). Though we cannot determine which one is a coincidence event, we can calculate the coincidence probability according to the matched radius. If the matched radius is larger, the coincidence probability of background galaxies is higher. In the H-ATLAS SDP field, the number density of galaxies with SDSS spectra ($r < 17.8$) is 102 per $\deg^2$, and that of *Herschel* 250 μm sources is approximately 484 per $\deg^2$. Adopting the $1\sigma$ positional errors of $2.4''$ (Smith et al. 2011), we obtain the coincidence probability of less than one object in 14 $\deg^2$. Although the probability is low, it is still considerable if the matched radius is enlarged to $3\sigma$. As the lensing system will magnify the background FIR/submm flux, it would also enhance the coincidence probability. Until now, at least two lensing systems in H-ATLAS (e.g. SDP.81 and SDP.130) have been discovered (e.g. Negrello et al. 2010; Frayer et al. 2011). SDP.81 is a background submillimeter galaxy which is lensed by an early-type of foreground galaxy.

## 5 SUMMARY

In this paper, we have studied a sample of 297 nearby galaxies which are cross-identified by *Herschel* and SDSS-DR7. We performed the morphological classification of these galaxies and studied their FIR/submm properties. Our main results are given below:

(1) We obtain a morphological fraction of E/S0:S(Sa-Sd):peculiar:compact = 0.08:0.38:0.34:0.08, the sample is inclined toward late-type galaxies, and more than 40% are peculiar/compact galaxies, which are related to interacting/merging systems.
(2) The peculiar/compact galaxies show higher FIR/submm luminosity-to-mass ratios than other types. The median ratios decrease from 100 μm to 500 μm. The sSFR derived from H$\alpha$ luminosity-to-mass ratio decreases from $\log \mathrm{sSFR} = -9.20$ for peculiar/compact galaxies to –9.88 for E/S0 galaxies.
(3) The correlations among five *Herschel* bands show that the correlation tends to be weaker with increasing wavelength. The Spearman rank coefficient changes from 0.98 in 160 μm to 100 μm to 0.88 in 500 μm to 100 μm, and the scatter from 0.09 dex to 0.25 dex follows an opposite trend. In the $L_{\mathrm{FIR/submm}}$-to-$L_{\mathrm{H}\alpha}$ correlations of star-forming galaxies, the 100 μm band presents the tightest correlation with H$\alpha$. As the wavelength increases, the Spearman rank coefficient decreases and the scatter increases, demonstrating the weak correlation between submm (*Herschel* SPIRE) bands and star formation. Therefore, *Herschel* SPIRE bands may not be good star formation tracers.

(4) The early-type galaxies deviate from the correlation of star-forming galaxies. The low luminosity AGNs do not show any difference from the star-forming galaxies in $L_{\rm FIR/submm}$-to-$L_{\rm H\alpha}$ correlations.

**Acknowledgements** We gratefully thank the anonymous referee for the constructive comments. We gratefully acknowledge S. Mao, F.S. Liu, X.M. Meng and C. Cao for their helpful discussions. This work is supported by the Key Laboratory of Optical Astronomy, National Astronomical Observatories, Chinese Academy of Sciences. This project is supported by the National Natural Science Foundation of China (Grant Nos. 11173030, 10833006, 10773014 and 11078017) and partly supported by the China Ministry of Science and Technology under State Key Development Program for Basic Research (2012CB821800).

*Herschel* is an ESA space observatory with scientific instruments provided by European-led Principal Investigator consortia and with important participation from NASA. H-ATLAS is a project associated with *Herschel* and its website is *http://www.h-atlas.org*. Funding for the creation and distribution of the SDSS archive has been provided by the Alfred P. Sloan Foundation, the Participating Institutions, the National Aeronautics and Space Administration, the National Science Foundation, the U.S. Department of Energy, the Japanese Monbukagakusho, and the Max Plank Society. The SDSS website is *http://www.sdss.org*. The SDSS is managed by the Astrophysical Research Consortium (ARC) for the Participating Institutions.